\documentclass[reqno,11pt]{amsart}
\usepackage{amsmath, latexsym, amsfonts, amssymb, amsthm, amscd,color}
\usepackage{amsmath, amsfonts, amssymb, amsthm, amscd, graphics,epsf,psfrag}

\advance\hoffset -.75cm

\numberwithin{equation}{section}

\makeatletter
\def\captionfont@{\footnotesize}
\def\captionheadfont@{\scshape}

\long\def\@makecaption#1#2{%
  \vspace{2mm}
  \setbox\@tempboxa\vbox{\color@setgroup
    \advance\hsize-6pc\noindent
    \captionfont@\captionheadfont@#1\@xp\@ifnotempty\@xp
        {\@cdr#2\@nil}{.\captionfont@\upshape\enspace#2}%
    \unskip\kern-6pc\par
    \global\setbox\@ne\lastbox\color@endgroup}%
  \ifhbox\@ne 
    \setbox\@ne\hbox{\unhbox\@ne\unskip\unskip\unpenalty\unkern}%
  \fi
  \ifdim\wd\@tempboxa=\z@ 
    \setbox\@ne\hbox to\columnwidth{\hss\kern-6pc\box\@ne\hss}%
  \else 
    \setbox\@ne\vbox{\unvbox\@tempboxa\parskip\z@skip
        \noindent\unhbox\@ne\advance\hsize-6pc\par}%
\fi
  \ifnum\@tempcnta<64 
    \addvspace\abovecaptionskip
    \moveright 3pc\box\@ne
  \else 
    \moveright 3pc\box\@ne
    \nobreak
    \vskip\belowcaptionskip 
  \fi
\relax
}
\makeatother
\def\writefig#1 #2 #3 {\rlap{\kern #1 truecm
\raise #2 truecm \hbox{#3}}}

\newtheorem{theo}{Theorem}[section]

\newtheorem{rem}[theo]{Remark}

\DeclareMathSymbol{\leqslant}{\mathalpha}{AMSa}{"36} 
\DeclareMathSymbol{\geqslant}{\mathalpha}{AMSa}{"3E} 
\DeclareMathSymbol{\eset}{\mathalpha}{AMSb}{"3F}     
\renewcommand{\leq}{\;\leqslant\;}                   
\renewcommand{\geq}{\;\geqslant\;}                   

\newcommand{\bra}{\langle}
\newcommand{\ket}{\rangle}

\newcommand{\cA}{\ensuremath{\mathcal A}}
\newcommand{\cB}{\ensuremath{\mathcal B}}

\newcommand{\cE}{\ensuremath{\mathcal E}}

\newcommand{\cI}{\ensuremath{\mathcal I}}

\newcommand{\cS}{\ensuremath{\mathcal S}}

\newcommand{\bbQ}{{\ensuremath{\mathbb Q}} }
\newcommand{\bbR}{{\ensuremath{\mathbb R}} }

\newcommand{\ga}{\alpha}
\newcommand{\gb}{\beta}
\newcommand{\gga}{\gamma}            
\newcommand{\gd}{\delta}

\newcommand{\gr}{\rho}

\newcommand{\gG}{\Gamma}
\newcommand{\gP}{\Phi}
\newcommand{\gD}{\Delta}

\newcommand{\gl}{\lambda}
\newcommand{\gL}{\Lambda}

\newcommand{\ex}{{\vec{e}_1}}

\newcommand{\jj}{J}

\newcommand{\haut}{{h}} 
\newcommand{\CA}{{\Omega}} 
\newcommand{\mean}{{\rm Mean}} 
\newcommand{\var}{{\rm Var}} 

\begin{document}
\title{Vortices in the two-dimensional Simple Exclusion Process}

\author{T. Bodineau$^{(1)}$}
\author{B. Derrida$^{(2)}$}
\author{Joel L. Lebowitz$^{(3)}$}

\address{
(1) D\'epartement de math\'ematiques et applications, Ecole Normale
Sup{\'e}rieure, CNRS-UMR 8553, 75230 Paris cedex 05, France}
\address{(2) Laboratoire de Physique Statistique, Ecole Normale
Sup{\'e}rieure, 24 rue Lhomond, 75231 Paris Cedex 05, France}
\address{(3) Department of Mathematics, Hill Center, Rutgers University, 110 Frelinghuysen Road, Piscataway, NJ 08854}

\date{18 January 2008}

\maketitle 

\begin{abstract} 

We show that the fluctuations of the partial current in two dimensional diffusive systems are dominated by vortices leading to
a different scaling from the one predicted by the hydrodynamic large deviation theory.
This is supported by exact computations of the variance of partial current fluctuations 
for the symmetric simple exclusion process on general graphs. 
On a two-dimensional torus, our exact expressions are compared to the results of numerical simulations. 
They confirm the logarithmic dependence  on the system size of the fluctuations of the partial flux.
The impact of the vortices on the validity of the fluctuation relation for partial currents is also discussed
in an Appendix.
\end{abstract}

\section{Introduction}
\label{sec: introduction}

Recently, it has  been shown how to compute the large deviation
function of the current in one dimensional diffusive
systems \cite{BDGJL4}-\cite{BD4}.  The hydrodynamic large deviation
theory \cite{BDGJL4,KOV,DV}, yields explicit expressions for the large 
deviation
function as well as the cumulants of the current fluctuations (under 
some stability condition \cite{BD1,BD4}).
The same hydrodynamical approach applies in principle also to 
currents in higher dimension.  In the present paper we show however 
that this approach does not always catch the correct scaling of the 
large deviations or of the cumulants of the current in higher dimensions.
This will be made explicit in the case of the 2 dimensional symmetric 
simple exclusion process (SSEP).

\bigskip

For a {\it one dimensional} diffusive system of length $L$ in contact at its
left end with a reservoir at density $\rho_a$ and  at its right end
with a reservoir  at density $\rho_b$, one can consider the total net
number $Q(\tau)$ of particles leaving the left reservoir during a time
interval  $\tau$.  This number $Q(\tau)$ fluctuates in time and one
expects that in the long time limit
\begin{equation}
{\rm Pro} \left( \frac{Q(\tau)}  {\tau} \simeq \jj \right) \sim \exp \big[- \tau G_L( \jj;\rho_a,\rho_b) \big]
\label{eq: 1}
\end{equation} 
where  $G_L(\jj ; \rho_a,\rho_b)$  is the large deviation function of
the flux through the system.
In fact $G_L$ does not depend on where the flux, i.e. the integrated
current, is measured along the one dimensional system, as long as
particles cannot accumulate.  For large $L$ and $\jj$ 
of order ${1 \over L}$, $G_L$ satisfies,  the following scaling
\begin{equation}
G_L(\jj;\rho_a,\rho_b) \simeq {1 \over L} F(L \jj; \rho_a,\rho_b)
\label{eq: 1.2}
\end{equation}

The scaling (\ref{eq: 1.2}) implies that for large $L$ all the 
cumulants of $Q(\tau)$ are of order $1/L$, i.e.
\begin{equation}
\lim_{\tau \to \infty}{\langle Q(\tau)^n \rangle_c \over \tau}  
\simeq {1 \over L} \kappa_n(\rho_a,\rho_b) \ .
\label{cum1}
\end{equation}
Explicit expressions  of the $\kappa_n(\rho_a,\rho_b)$ have been
obtained \cite{BD1,BD4}
in terms of 
the diffusion constant $D(\rho)$ 
and the conductivity $\sigma(\rho)$ \cite{spohn}.
One can also show that the large deviation function $G_L$ of the 
current satistifies the fluctuation theorem 
\cite{ECM,GC,K,LS,M1,ES,G2,BD1,BDGJL4,BD4}, i.e.
\begin{equation}
G_L(\jj;\rho_a,\rho_b) - G_L(-\jj;\rho_a,\rho_b)= \jj [ \log z(\rho_b) - \log z(\rho_a)]
\label{ft}
\end {equation}
where $z(\rho)$ is the fugacity of a  reservoir at density $\rho$.

\begin{figure}[h]
\begin{center}
\leavevmode
\epsfysize = 5 cm
\epsfbox{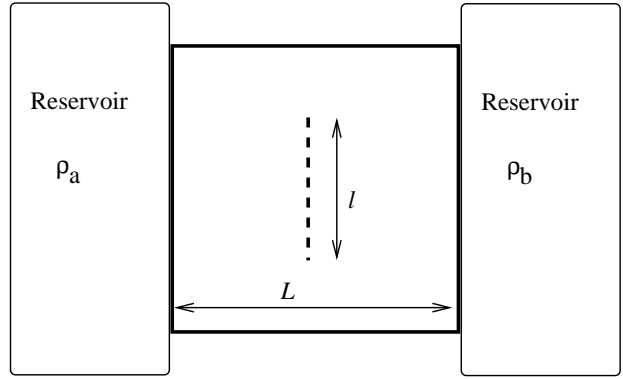}
\end{center}
\caption{We are going to consider the distribution of the current flowing through the dashed vertical slit of length  $\ell < L$.}
\label{fig: reservoirs}
\end{figure}

\bigskip

In {\it higher  dimension},  one can study, as in one dimension, the
total current flowing through the system  from one reservoir to the
other, but one can also study part of this current.
In this paper, we consider the SSEP on a square
lattice of size $L$, with periodic boundary conditions in the vertical
direction and study the current flowing through
a  vertical slit of length  $\ell < L$ (see figure \ref{fig: reservoirs}).
The large deviation function $G_{L,\ell}( \jj ;\rho_a,\rho_b)$,
defined as in 
(\ref{eq: 1}), 
depends of course 
on the size $\ell$ of the slit.  
One reason for considering the fluctuations of this partial current is
that in experiments it is often only possible to measure the fluctuations of
local quantities and not of global quantities \cite{Ciliberto,BL}.

In two dimensions, when $\ell=L$, i.e. when one considers the total
current flowing through the system, the large deviation function
derived from the hydrodynamic theory satisfies for large $L$ and $\jj$
of order 1 a scaling similar to the one dimensional case \cite{BDGJL6}
\begin{equation}
G_{L,L}(\jj;\rho_a,\rho_b) \simeq F(\jj ; \rho_a,\rho_b)
\label{eq: LD 2D}
\end{equation}
(this would become $L^{d-2} F(L^{2-d}\jj;\rho_a,\rho_b)$ for a cube of
size $L$ in dimension $d$ and $\jj$ of order $L^{d-2}$).
In the present paper we show that $G_{L,\ell}$ cannot satisfy the same scaling (\ref{eq: LD 2D})
as $G_{L,L}$ and that  
for large $L$, if one keeps the ratio $h=\ell/L$ fixed, then for all $0<h<1$ and $\jj$ of order 1
\begin{equation}
G_{L,Lh}(\jj;\rho_a,\rho_b)  \to 0  {\rm \ \ \ as   \ \ \ }  L \to \infty \ .
\label{G0}
\end{equation}

While, as in (\ref{cum1}), one expects the cumulants of the total flux
$Q(\tau)$ to have a large $L$ limit  
\begin{equation}
\lim_{\tau \to \infty}{\langle Q(\tau)^n \rangle_c  \over \tau}\to   \kappa_n(\rho_a,\rho_b) \ .
\end{equation}
(which would become $\frac{1}{\tau} \langle Q(\tau)^n \rangle_c \simeq L^{d-2}
\kappa_n(\rho_a,\rho_b)$ in dimension $d$), we will see  by an explicit calculation of the second cumulant that for $\ell= L h$, 
\begin{equation}
\lim_{\tau \to \infty}{\langle Q^{(h)}(\tau)^2 \rangle_c  \over \tau}
\sim \log L {\rm  \ \ \ as   \ \ \ }  L \to \infty,  {\rm  \ \ \ when   \ \ \ } \  0 < h < 1 \, ,
\label{eq: log scale}
\end{equation}
where $Q^{(h)}(\tau)$ is the flux of particles through the slit during time $\tau$.

The fluctuation theorem, which is satisfied as written in (\ref{ft})  for
the two-dimensional SSEP when $\jj$ is  the total current through the 
system (i.e. when $\ell=L$), has in fact no reason to remain valid for 
$\ell<L$: in the large $L$ limit, the difference $G_{L,L \haut}(\jj) - G_{L,L \haut}(- \jj)$ 
vanishes when $0<h<1$ so that (\ref{ft}) cannot hold and 
a singular dependence can be expected in $G_{L,L \haut} (J)$ when $h \to 1$.
In Appendix A, we give a simple example of a two site model where one
can see clearly that the fluctuation theorem is satisfied when one
looks at the total current but is no longer valid when one considers 
only part of the current see \cite{gaspard} for a discussion on the 
validity of the fluctuation theorem for partial currents.

\bigskip

The rest of the paper is organized as follows. In section \ref{sec:
 vortices}, we recall the hydrodynamic large deviation theory
 \cite{BDGJL4,BD2} and show the asymptotics (\ref{G0}).
Although the hydrodynamic large deviation theory does not predict the 
correct scaling of the current deviation, the analysis of section 
\ref{sec: vortices} suggests that local current fluctuations are
 dominated by vortices. 
Restarting at the microscopic level, the variance of the integrated current is
 computed for the SSEP on a general graph (section \ref{sec: General Graph}) and explicit expressions
 are obtained for the current through a slit for the SSEP on a two-dimensional torus (section \ref{sec: Z2}).
Our exact expression leads to the asymptotics of the form  (\ref{eq: log scale}) and are compared to the 
results of numerical simulations. 
Finally the appendices are devoted to comments on the 
fluctuation relation (\ref{ft}) for partial currents, and to some technical calculations.  
We note that sections \ref{sec: vortices}, \ref{sec: General Graph} and Appendix A can be read independently.

\section{Vortices and current fluctuations}
\label{sec: vortices}

For simplicity, we briefly recall the large deviation hydrodynamic
limit theory in the framework of the two-dimensional SSEP on the
square lattice in the periodic domain 
$\gL = \{1,L\}^2$. At the microscopic level, 
each particle jumps randomly with rate 1 to
a nearest neighboring site and the jump is allowed only if the
neighboring site is empty.  After rescaling space by $1/L$ and time 
by $1/L^2$, the macroscopic density $\gr (x,t)$ obeys the diffusion
equation \cite{spohn,KL} in the  macroscopic domain 
$\widehat \gL = [0,1]^2$, (with periodic boundary conditions),
\begin{eqnarray}
\label{eq: limit hydro}
\partial_t \gr (x,t) =  \Delta \gr (x,t), \quad   x = (x_1,x_2) \in \widehat \gL, 
\end{eqnarray}
where $\Delta$ denotes the Laplacian. 
One can also define a macroscopic current $j(x,t) = (j_1(x,t), j_2(x,t))$ in the 
directions $\vec{e}_1, \vec{e}_2$ which has to satisfy
\begin{eqnarray*}
\partial_t \gr(x,t) = - \nabla \cdot j(x,t) \, .
\end{eqnarray*}
The rescaled current $j$ is such that if $q_{(i,i+\vec{e}_\ga)}(\tau)$ 
is the microscopic integrated current through the bond
$(i,i+\vec{e}_\ga)$ (with $\ga =1$ or $2$) over the microscopic time 
interval [0, $\tau$], then for a system of size $L$ and times $\tau$ of
order $L^2$, 
one has $q_{(i,i+\vec{e}_\ga)}(\tau) = L \int_0^{\tau/L^2} j_\ga 
(\frac{i}{L},t) \, dt$.

\bigskip

Using the hydrodynamic large deviation theory, we are going to show
that the scaling of the large deviations is different for the current flowing 
through the whole system or through a slit 
(as in figure \ref{fig: reservoirs}).

\subsection{Total current deviations}

We denote by $Q(\tau)$ the integrated total current during the 
microscopic time interval $[0,\tau]$ through a vertical section of the 
whole system, say the current flowing through the edges 
$\{ (L/2,i_2), (L/2 + 1,i_2)\}_{1\leq i_2 \leq L}$. 
The corresponding large deviation function $G_{L,L}$  is defined by
\begin{eqnarray}
\label{eq: hydro scale}
\lim_{\tau \to \infty}
\; -{1 \over \tau} \log {\rm Pro} 
\left( \frac{Q(\tau)}{\tau} \approx J \right) = G_{L,L} (J)   \, ,
\end{eqnarray}
where ${\rm Pro} \left( \frac{Q(\tau)}{\tau} \approx J \right)$
denotes the probability of observing a total current $J$ in the 
$\vec{e}_1$ direction averaged over the microscopic time interval $[0,\tau]$.
According to the large deviation hydrodynamic theory, one expects, in
accord with (\ref{eq: LD 2D}) that 
$\lim_{L \to \infty} G_{L,L}(J) = F(J)$  where the function $F(J) 
= \lim_{T \to \infty} F_T (J)$ with
\begin{eqnarray}
\label{eq: mother}
F_T(J)=  \inf_{j,\gr} 
\left\{ {1 \over T} \; \cI_T (j,\gr)  \right\} \, ,
\quad \text{and} \quad 
\cI_T (j,\gr) =  \frac{1}{2}  \int_0^T  \int_{\widehat \gL} \, dt \,
dx \ {\big( j_1 + \partial_{x_1} \gr  \big)^2 + \big( j_2 + \partial_{x_2} 
\gr  \big)^2 \over 2 \gr (1-\gr)}.
\end{eqnarray}
The minimum is taken over the macroscopic evolutions $\{ \gr(x,t),
j(x,t) \}$ during the macroscopic time interval $[0,T]$ which satisfy
the constraints
\begin{eqnarray}
\label{eq: courant total} 
\partial_t \gr(x,t) = - \nabla \cdot j(x,t),
\qquad \text{and} \qquad
J = \frac{1}{T} \int_0^T \int_0^1 \, dt \, dx_2 \  j_1 \left( \left({1 \over 2}, x_2\right),t \right)  \, .
\end{eqnarray}

\begin{rem}
Note that the mathematical statement from the hydrodynamic limit theory \cite{BDGJL5}
relies on a more involved asymptotic with a joint space/time scaling:
instead of (\ref{eq: hydro scale}), the large deviation function for a total 
current $J$ over the microscopic time interval $[0,L^2 T]$ is given  by  
\begin{eqnarray*}
\lim_{L \to \infty} \  -{1 \over L^2 T} \log {\rm Pro}
\left( \frac{Q(L^2 T)}{L^2 T}  \approx J \right) = F_T (J)   \, ,
\end{eqnarray*}
where $F_T$ has been introduced in (\ref{eq: mother}).
When writing (\ref{eq: hydro scale}), (\ref{eq: mother}), we assumed that in the 
previous expression the limits $L \to \infty$ and $T \to \infty$ can be exchanged.
\label{rem: 1}
\end{rem}

\bigskip

As we consider in this section a system with periodic boundary
conditions and no sources,
the steady state is the equilibrium one in which all configurations with
a specified total of number particles have equal weight. The mean
current through the system is therefore 0 and we 
are going to show that for any current deviation $J \not = 0$
\begin{eqnarray}
\label{eq: G >0}
F(J)>0 \,  .
\end{eqnarray}
Expanding $\cI_T$ in \eqref{eq: mother} and using Jensen's inequality leads to 
\begin{eqnarray}
\label{eq: 2.6}
&& \cI_T (j,\gr) =  \frac{1}{2}  \int_0^T  \int_{\widehat \gL} \, dt \, dx \ 
\left[ {(j_1 )^2 + (j_2)^2 \over 2 \gr(1-\gr)} + {\big(  \nabla \gr
\big)^2 \over 2 \gr(1-\gr)} \right] + C_T        \qquad \\
&& \qquad \qquad 
 \geq   \int_0^T  \int_{\widehat \gL}  \, dt \, dx \  {(j_1 )^2  } + C_T 
\geq   T \left( {1 \over T} \int_0^T  \int_{\widehat \gL} \, dt \, dx \ j_1 \right)^2 + C_T  \, ,\nonumber
\end{eqnarray}
where $C_T$ is the contribution of the cross terms in \eqref{eq: mother}
$$ C_T = \frac{1}{2} \int_0^T  \int_{\widehat \gL} \, dt \, dx \ \frac{j \cdot \nabla \gr}{\gr (1-\gr)}
= \frac{1}{2} \int_{\widehat \gL} \, dx  \left\{  \cS(\gr(x,0)) - \cS(\gr(x,T)) \right\} $$
with $\cS(\gr) = - [\gr \log(\gr) + (1 - \gr) \log(1- \gr)]$.
As it is equivalent to measure the total current through any section of the system,
the constraint \eqref{eq: courant total} on the current deviations becomes
\begin{eqnarray}
\label{eq: 2.7}
J = \frac{1}{T} \int_0^T \int_{\widehat \gL} \, dt \, dx \  j_1 \left(
x, t \right)  .
\end{eqnarray}
Thus  $\inf_{j,\gr} \big\{ \cI_T (j,\gr) \big\} \geq T J^2 + C_T$.
As $C_T$ remains bounded in time, \eqref{eq: G >0} follows from (\ref{eq: mother}) .

	\subsection{Partial current deviations}

The functional $\cI_T$ defined in (\ref{eq: mother}) should in principle provide the large deviations of the current
through any macroscopic region of the system. 
We consider now a slit of macroscopic height $\haut<1$ (the segment
$[(1/2,0),(1/2,\haut)])$ and denote by $Q^{(h)}(\tau)$ the integrated current through the slit during the microscopic time interval $[0,\tau]$, i.e. the current flowing through the edges $\{ (L/2,i_2), (L/2 + 1,i_2)\}_{1\leq i_2 \leq \haut L}$.
Then, the large deviation function for observing a current deviation $J \not = 0$ is given by 
\begin{eqnarray*}
\lim_{\tau \to \infty}
\; -{1 \over \tau} \log \; {\rm Pro} \left( \frac{Q^{(h)} (\tau)}{\tau} \approx J \right) = G_{L,Lh} (J) \, .
\end{eqnarray*}
One expects from (\ref{eq: LD 2D}), that $\lim_{L \to \infty} G_{L,Lh}(J) = F_h(J)$  with
\begin{eqnarray}
\label{eq: zero func}
F_h (J) = \lim_{T \to \infty} \inf_{j,\gr} \ \left\{{ 1\over T}  \cI_T
(j,\gr) \right\}  
\end{eqnarray}
where $\cI_T (j,\gr)$ is defined in (\ref{eq: mother}) and the macroscopic constraints (\ref{eq: courant total})
are replaced by
\begin{eqnarray}
\label{eq: 2.9}
\partial_t \gr(x,t) = - \nabla \cdot j(x,t),
\qquad \text{and} \qquad
J = {1 \over T}  \int_0^T \, \int_0^\haut \, dt \, dx_2 \ \  j_1
\left( ({1 \over 2}, x_2),t \right) \, .
\end{eqnarray}
We are going to show that in contrast to \eqref{eq: G >0}, the large deviation
function $F_h$ in \eqref{eq: zero func} vanishes for $0< \haut <1$ 
(as claimed in (\ref{G0})).

\bigskip

One can bound (\ref{eq: zero func}) by
\begin{eqnarray*}
\inf_{j,\gr} \left\{ \frac{1}{T} \; \cI_T (j,\gr)  \right\} \leq \tilde F_h(J) \, ,
\end{eqnarray*}
where the functional $\tilde F_h (J)$ is the restriction of $\cI_T$ to time independent  density and current profiles,
\begin{eqnarray}
\label{eq: mother 2}
\tilde F_h (J)=  \inf_{j,\gr} \left\{  \frac{1}{2} \int_{\widehat \gL} \, dx 
\left[ {(j_1 )^2 + (j_2)^2 \over 2 \gr(1-\gr)} + {\big(  \nabla \gr \big)^2 \over 2 \gr(1-\gr)} \right] 
\right\} 
\end{eqnarray}
where the density and current constraints satisfy
\begin{eqnarray}
\label{eq: statio current}
0 = \nabla \cdot j =  \partial_1 j_1 (x) +  \partial_2 j_2 (x),
\qquad \text{and} \qquad
J = \int_0^\haut \, dx_2 \ \  j_1 \left( {1 \over 2}, x_2 \right)  \, .
\end{eqnarray}

\begin{figure}[h]
\begin{center}
\leavevmode
\epsfysize = 4 cm
\epsfbox{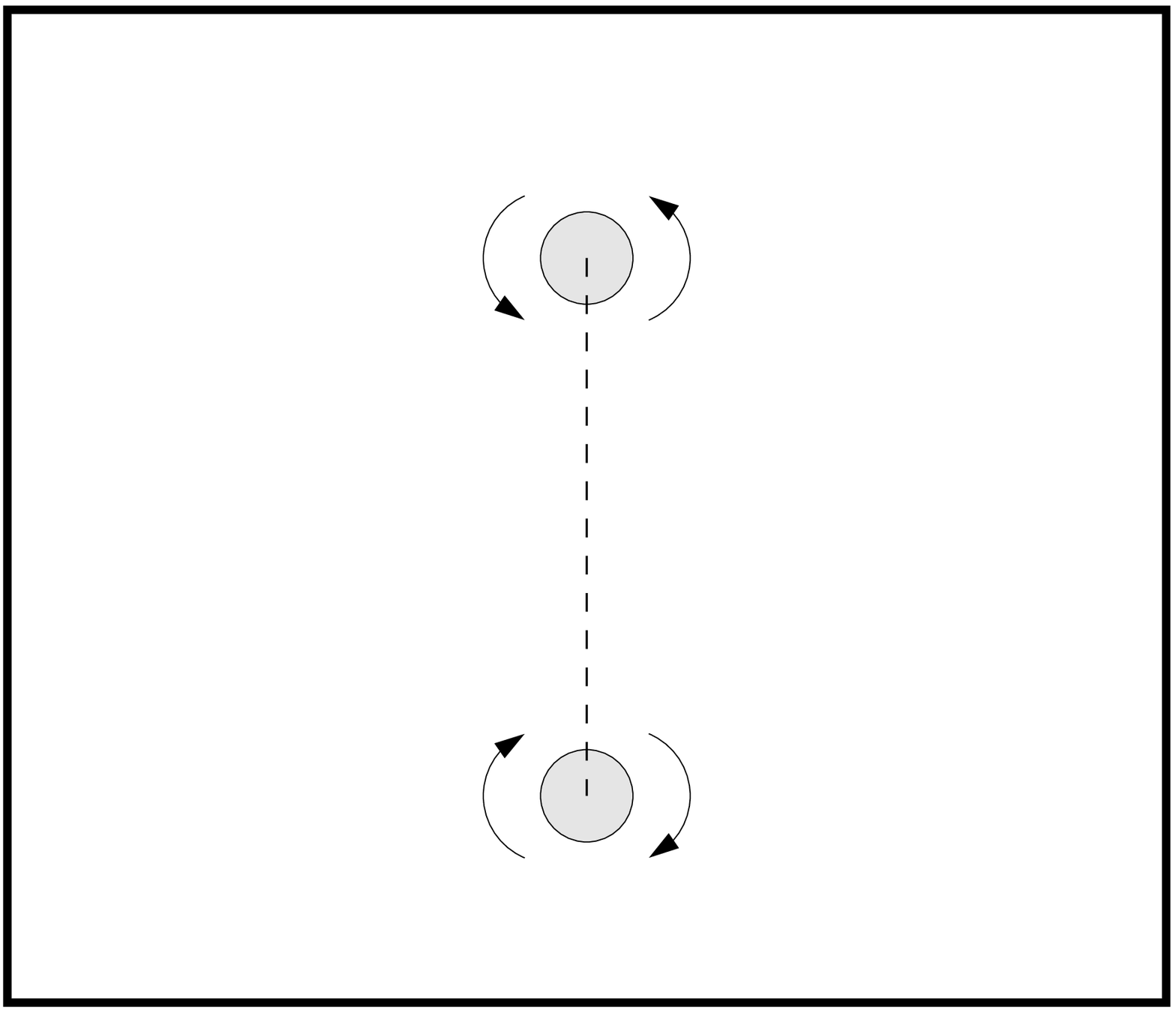} \hskip3cm
\epsfysize = 4 cm
\psfrag{a}[cr]{$r_2$}
\psfrag{b}[cc]{$r_1$}
\epsfbox{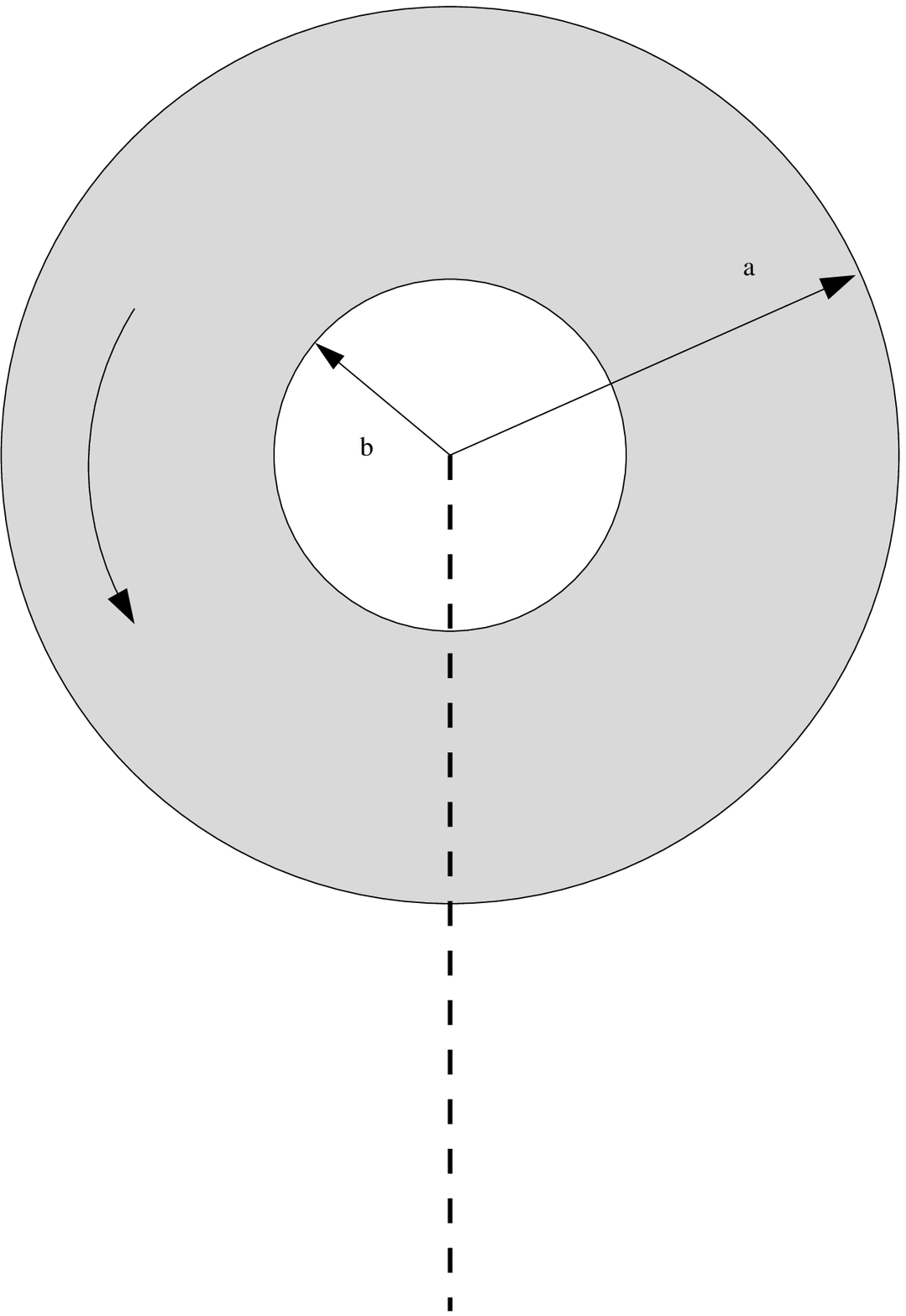}
\end{center}
\caption{On the left, two vortices located at the edges of the dashed slit are depicted. On the right, a blow up of one vortex concentrated on the disk of radius $(r_1,r_2)$: the current (\ref{eq: test current}), 
(\ref{eq: local vortex}) is proportional to $1/r$ for $r$ in $(r_1,r_2)$ and vanishes outside this annulus.}
\label{fig: vortices}
\end{figure}

A guess to bound (\ref{eq: mother 2}) is to consider the equilibrium density (uniformly equal to the constant 
density $\bar \gr$) and a current deviation consisting of two vortices localized at the edges of the slit 
$(1/2,0)$ and $(1/2,\haut)$ (see figure \ref{fig: vortices})
\begin{eqnarray}
\label{eq: test current}
\forall x \in \widehat \gL, \qquad 
j(x) =  J \Big(  \gP \big(x - (1/2,\haut) \big)- \gP \big( x - (1/2,0) \big)  \Big) \, ,
\end{eqnarray} 
where $\gP$ denotes the vector
\begin{eqnarray}
\label{eq: local vortex}
\forall x = (x_1,x_2), \qquad 
\gP \big( x \big)=  {1 \over 2 \log (r_2/r_1)} \;  
{ 1_{ \{r_1 \leq \sqrt{x_1^2 + x_2^2} \leq r_2 \}} \over x_1^2 + x_2^2} \; \big(-x_2,x_1 \big)  \, ,
\end{eqnarray} 
with $r_1<r_2\ll 1$. One can check that the current defined in 
(\ref{eq: test current}) satisfies the constraint (\ref{eq: statio
current}).  Furthermore, we can bound $\tilde F_h (J)$ by, 
\begin{eqnarray}
\label{eq: loga}
\tilde F_h (J) \leq   \frac{1}{4 \bar \gr ( 1- \bar \gr)} \int_{\widehat \gL} \, dx \left[(j_1 )^2 + (j_2)^2  \right] 
= \frac{\pi}{4 \bar \gr ( 1- \bar \gr)} \; \frac{J^2}{\log(r_2 / r_1)}  \, .
\end{eqnarray}
Letting $r_1,r_2$ go to 0 while ${r_2 \over r_1}$ diverges, we find that $\tilde  F_h (J) =0$ so that 
the large deviation cost in (\ref{eq: zero func}) is 0.

\begin{rem}
On a finite lattice of size $L$, the current has to flow through the bonds and therefore the 
ratio ${r_2 / r_1}$ is at most $L$. This imposes a cut-off and the computation (\ref{eq: loga}) based 
on (\ref{eq: mother 2}) leads to $G_{L,Lh} (J)$ of order ${1 \over \log L}$.
This logarithmic dependence will be confirmed for the SSEP by a direct computation of the current fluctuations in sections \ref{sec: General Graph} and \ref{sec: Z2}.
\end{rem}

Expression (\ref{eq: loga})  shows that the cost of the fluctuations due to the vortices
is low and one may wonder if the vortices are the optimal minimizers of (\ref{eq: mother 2}) or whether one should expect a more complex structure.
To check this, we first note that the current $j$ in 
\eqref{eq: statio current} 
is divergence free, thus it can be represented as the sum of the curl of a vector field 
$\big( 0,0,\Psi(x_1,x_2) \big)$ 
and a constant vector field $(C_1,C_2)$
\begin{eqnarray}
\label{eq: stream}
j = \nabla \times (0,0,\Psi) + (C_1,C_2) = (\partial_2 \Psi, - \partial_1 \Psi) + (C_1,C_2)
\end{eqnarray}
and the current constraint (\ref{eq: statio current}) becomes
\begin{eqnarray}
\label{eq: stream current}
J = \Psi(1/2,\haut) -\Psi(1/2,0)  + C_1 \haut \, .
\end{eqnarray}
Choosing the density equal to $\bar \gr$, (\ref{eq: mother 2}) reduces to the varitional principle
\begin{eqnarray}
\label{eq: Psi}
\tilde F_h (J) \leq {1 \over 4 \bar \gr ( 1- \bar \gr)}  \; \inf_{\Psi,(C_1,C_2)} \;
\left\{ \int_{\widehat \gL} \, dx \; \big(\partial_1 \Psi \big)^2  + \big(\partial_2 \Psi \big)^2 + C_1^2 + C_2^2 \right\} \, .
\end{eqnarray}
where $\Psi$ and $C_1$ satisfy the constraint \eqref{eq: stream current}.
The solutions of the above variational problem will satisfy 
$$\Delta \Psi (x) = \ga \left( \gd_{x, (1/2 , h)} -  \gd_{x, (1/2 , 0)}   \right) \, ,$$ 
where $\ga$ is the Lagrange parameter associated to the constraint (\ref{eq: stream current}).
This would lead to  a $\Psi$ which diverges logarithmically at the edges of the slit 
and therefore cannot satisfy condition (\ref{eq: stream current}) for any non-zero $\ga$.
Nevertheless, using a cut-off similar to $r_1$ in (\ref{eq: local vortex}), we can recover the  vortex like structures (\ref{eq: test current}).

\begin{rem}
For more general diffusive systems the hydrodynamic large deviations are governed by functionals of the type
(\ref{eq: mother}) which depend on diffusion and conductivity matrices \cite{BDGJL4}.
One could extend the previous discussion to these cases and  the large deviation function $F_h(J)$ 
would vanish as soon as $h< 1$.
For open systems, similar computations can be done as the fluctuations are dominated by vortices 
localized at the edges of the slit (\ref{eq: test current}) and the reservoirs play a negligible role.
\end{rem}

\begin{rem}  We note that in analogy to (\ref{eq: 2.7})
we can consider the partial current specified in (\ref{eq: 2.9}) as a
limiting case of an integrated current in a domain $B\subset \widehat \gL,$
\begin{eqnarray}
\label{eq: 2.18}
J_B ={1\over T} \int_0^T \int_B j_1(x, t) dx  \, .
\end{eqnarray}
Taking $B$ to be a rectangle of height $h$ and width $w$ we get
that the flux through the line segment $h$ is given by $w^{-1}
J_B$, in the limit $w \to 0$.  For the large deviation of $J_B$ one
can repeat the analysis leading to (\ref{eq: 2.6}) yielding 
\begin{eqnarray}
\inf_{j,\gr} \cI_T(j, \rho:B)\geq T \frac{J_B^2}{|B|} + C_T  \, .
\end{eqnarray}
This non-zero lower bound reflects the fact that any vortex flow used to
implement the fluctuation $J_B$ will have to be of a size $w$
or greater.

\end{rem}

\section{Current fluctuations on a general Graph}
\label{sec: General Graph}

In this section, we consider the SSEP on a general connected graph $(\gL,\cE_\gL)$ where $\gL$ is a finite 
set of sites and $\cE_\gL$ the set of edges on which particles jump with rate 1 according to the exclusion rule.
We also suppose that particles are created and annihilated at the sites in the subset $\CA$ of $\gL$ ($\CA$ may be empty). For any site $i$ in $\CA$, we suppose that creation and annihilation occur at rate $\ga_i$ $\gb_i$ (and for simplicity we choose $\ga_i + \gb_i =1$).
In section \ref{sec: Z2}, we will apply the results obtained for general graphs to the microscopic domain $\gL = \{1,L\}^2$ and derive explicit expressions in this case.

If $(i,j)$ is an edge  in $\cE_\gL$ then the number of  particle jumps
from $i$ to $j$ during the time interval $\tau$ is denoted by $q_{(i,j)} (\tau)$. The current flowing through $(i,j)$ during time $\tau$ is then $q_{(i,j)} (\tau) - q_{(j,i)} (\tau)$.
If creation and annihilation occur,  we enlarge the graph $(\gL,\cE_\gL)$  by associating to each site $i$ in $\CA$ a new site $\bar i$. The site $\bar i$ can be interpreted as a source and we denote by $q_{(\bar i,i)} (\tau)$ the number of particles created at $i$ and $q_{(i,\bar i)} (\tau)$ the number of particles annihilated at site $i$.
It is convenient to enlarge the graph $\cE_\gL$ into $\bar \cE_\gL$ by adding to the original graph the new edges $(i,\bar i)_{i \in \CA}$ and 
$(\bar i, i)_{i \in \CA}$. 
We denote by $\bar \CA$ the set of the new sites, and by $\bar \gL$ the union of $\gL$ and $\bar \CA$.

\bigskip

For any field $\{\cA_b \}_b$ indexed by the edges and such that for any edge $(i,j)$,  $\cA_{(i,j)}=-\cA_{(j,i)}$,  
we are going to study the fluctuation  of the integrated current defined by
\begin{eqnarray}
\label{eq: bbQ}
\bbQ^\cA(\tau)  =\sum_{b \in \bar \cE_\gL} \cA_b \, q_b(\tau) \, ,
\end{eqnarray}
where the sum is over all the oriented edges $b$.
The field $\cA_b$ can be thought as a test function.

\bigskip

One  can define the divergence and the gradient on the graph.
For any field $\{\cA_b \}_b$ and any site $i$ in $\gL$
\begin{eqnarray}
\label{eq: div}
{\rm div} \cA (i) = \sum_{j \sim i} \cA_{(i,j)} \, ,
\end{eqnarray}
where the sum is over all the edges leaving site $i$ (this includes the edges  $(i,\bar i)$ if creation or annihilation occur at site $i$). 
For any function $H_i$ in $\bar \gL$, the gradient is a function indexed by the edges $b =(i,j)$ in $\bar \cE_\gL$
\begin{eqnarray}
\label{eq: grad}
\nabla_b H = H_j - H_i \, .
\end{eqnarray}
In the following, we will consider only functions $H$ in $\bar \gL$ equal to 0 in $\bar \CA$.

\subsection{Gauge invariance}

Before, computing the variance of $\bbQ^\cA (\tau)$ defined in (\ref{eq: bbQ}), we first show that 
for large  $\tau$ similar asymptotics of  $\log \bra \exp \big( \gl \bbQ^\cA (\tau) \big) \ket$ can 
be obtained for different choices of $\{\cA_b \}_b$.

\medskip

For any site $i$ in $\gL$, one has with notation (\ref{eq: div})
\begin{eqnarray}
\label{eq: current div =0}
\eta(i,\tau) - \eta(i,0)=  \sum_{j \in \bar \gL, \atop j \sim i} q_{(j,i)}(\tau)-q_{(i,j)}(\tau)  \, .
\end{eqnarray}
This implies that for any function $H$ on $\bar \gL$ (equal to 0 on $\bar \CA$)
\begin{eqnarray}
\label{eq: zero}
\bbQ^{\nabla H} (\tau)=
\sum_{b \in \bar \cE_\gL} \nabla_b  H \; q_b(\tau) = \sum_{i \in \bar \gL} H_i  \;  
\left(\sum_{j \in \bar \gL,\atop j \sim i} q_{(j,i)}(\tau)-q_{(i,j)}(\tau) \right)
=   \sum_{i \in \gL} H_i  \;  (\eta(i,\tau) - \eta(i,0)) \, .
\end{eqnarray}
Therefore, $\bbQ^{\nabla H} (\tau)$  remains bounded when the time diverges.
As a consequence, for any $\gl \in \bbR$, $\cA_b$ and $H_i$
\begin{eqnarray}
\label{eq: gauge}
\lim_{\tau \to \infty}  {1 \over \tau} \log
\left \bra \exp \left( \gl \sum_{b \in \bar \cE_\gL} \cA_b \, q_b(\tau) \right) \right \ket
=
\lim_{\tau \to \infty}  {1 \over \tau} \log
\left \bra \exp \left( \gl \sum_{b \in \bar \cE_\gL} (\cA_b + \nabla_b H) \, q_b(\tau) \right) \right \ket \, .
\end{eqnarray}
Thus the large deviations of $\bbQ^\cA (\tau)$ and $\bbQ^{\cA + \nabla H}(\tau)$ 
with respect to time are the same.

\medskip

We are going now to recall how a field $\cA_b$ in $\bar \cE_\gL$ can be decomposed as
\begin{eqnarray}
\label{eq: decomposition}
\cA_b= V_b + \nabla_b H \, ,
\end{eqnarray}
where $H$ is a function in $\bar \gL$ (equal to 0 on $\bar \CA$) and $V$ is divergence free in $\gL$
$$\forall i \in \gL, \qquad \qquad {\rm div} V(i) =0. $$
Note that no conditions are imposed on ${\rm div} V(\bar i)$ for $\bar i$ in $\bar \CA$, if there are sources ($\CA \not = \emptyset$).
If there are no sources ($\CA = \emptyset$), then $H$ is defined up to a constant.

For decomposition (\ref{eq: decomposition}) to hold, $H$ has to be the solution of
\begin{eqnarray}
\label{eq: laplacien}
\forall i \in \gL, \qquad {\rm div} \cA (i) = \Delta H_i = \sum_{j \sim i} (H_j - H_i) \, .
\end{eqnarray}
In the case with sources ($\CA \not = \emptyset$), the solution of 
\eqref{eq: laplacien} can be written in terms of the Green's
functions, defined for any site $k$ by 
\begin{eqnarray}
\label{eq: Green general}
\forall i \in \gL,\quad  \Delta G^{(k)}_i = - \gd_{i, k}, \qquad {\rm and} \qquad
\forall i \in \bar \CA, \quad G^{(k)}_i =0 \, .
\end{eqnarray}
Thus for $j \in \bar \gL$
\begin{eqnarray}
\label{eq: H}
H_j = - \sum_{k \in \gL}  {\rm div} \cA(k) \; G^{(k)}_j \, ,
\end{eqnarray}
and the field $V_b = \cA_b - \nabla_b H$  is divergence free.

\subsection{Variance of the current}

We are  going to compute the variance of 
$\bbQ^\cA(\tau)  =\sum_{b \in \cE_\gL} \cA_b \, q_b(\tau)$.
From (\ref{eq: gauge}--\ref{eq: decomposition}), we know that 
to compute large time asymptotics it is enough to consider $\cA$
which is divergence free.

\bigskip

One has
\begin{eqnarray}
&& \partial_\tau  \log \left \bra \exp \left( \gl \bbQ^\cA(\tau)  \right) \right \ket
= \nonumber \\
&& \quad \sum_{i \in \CA} 
\ga_i \big( \exp(  \gl \cA_{(\bar i,i)}) -1 \big) \; 
{\left \bra (1-\eta_i)  \exp \left( \gl \bbQ^\cA(\tau)   \right) \right \ket \over \left \bra \exp \left( \gl \bbQ^\cA(\tau)   \right) \right \ket}
 + \gb_i \big( \exp( - \gl \cA_{(\bar i,i)}) -1 \big) 
{\left \bra \eta_i  \exp \left( \gl \bbQ^\cA(\tau) \right) \right \ket \over \left \bra \exp \left( \gl \bbQ^\cA(\tau)   \right) \right \ket} \nonumber  \\
&& \quad + \sum_{(i,j) \in \cE_\gL} 
(\exp( \gl \cA_{(i,j)}) - 1) {\left \bra \eta_i (1- \eta_j) \exp \left( \gl \bbQ^\cA (\tau) \right) \right \ket \over \left \bra \exp \left( \gl \bbQ^\cA(\tau)   \right) \right \ket} \, ,
\label{eq: derivee}
\end{eqnarray}
where the sum is over all the oriented bonds $(i,j)$ and $\bra \cdot \ket$ denotes the average over the random process in the time interval $[0,\tau]$ and over an initial condition chosen according to the invariant measure for the SSEP.
The procedure to derive \eqref{eq: derivee} is similar to what was done in \cite{DDR}.
One considers all the possible moves occurring during an infinitesimal time interval $d \tau$ and their contributions to $ \left \bra \exp \left( \gl \bbQ^\cA(\tau)  \right) \right \ket$. The first terms in \eqref{eq: derivee} correspond to a jump of a particle from site $\bar i$ to site $i$ (creation) or from 
site $i$  to site $\bar i$ (annihilation), whereas the last term corresponds to a jump from site $i$ to $j$.

Let us denote by $\bra \cdot \ket_\gl$ the expectation of the tilted measure: for any function $f$
\begin{eqnarray}
\label{eq: tilted measure}
\bra f(\eta) \ket_\gl = \lim_{\tau \to \infty}
{\left \bra f\big( \eta(\tau) \big) \exp \left( \gl  \bbQ^\cA(\tau)  \right) \right \ket \over \left \bra \exp \left( \gl \bbQ^\cA(\tau)   \right) \right \ket}\, .
\end{eqnarray}

Using the symmetry $\cA_{(i,j)} = - \cA_{(j,i)}$ and the relation $\ga_i+\gb_i= 1$, we get by
expanding (\ref{eq: derivee}) for small $\gl$  
\begin{eqnarray}
\label{eq: derivee 2}
&&  \lim_{\tau \to \infty} \partial_\tau  \log \left \bra \exp \left( \gl  \bbQ^\cA (\tau) \right) \right \ket
= \gl \sum_{i \in \CA} \cA_{(\bar i,i)}  \ga_i  
+  \gl \; \sum_{i \in \gL}  {\rm div} \cA (i) \; \left \bra \eta_i \right \ket_\gl \\ 
&& \qquad \qquad 
+ {\gl^2 \over 2}  \sum_{i \in \CA} (\cA_{(\bar i,i)})^2  
\left \bra \big( \gb_i \eta_i + \ga_i (1 - \eta_i) \big)\right \ket_\gl 
+ {\gl^2 \over 2}  \sum_{(i,j) \in \cE_\gL} (\cA_{(i,j)})^2  \left \bra \eta_i (1- \eta_j)  \right \ket_\gl  +  O(  {\gl^3}) \, . \nonumber 
\end{eqnarray}
For $\gl$ small, one expects that 
\begin{eqnarray}
\label{eq: corrections lambda}
\left \bra \eta_i \right \ket_\gl  = \left \bra \eta_i \right \ket + O(\gl),
\qquad 
\left \bra \eta_i (1- \eta_j)\right \ket_\gl  = \left \bra \eta_i (1- \eta_j)\right \ket + O(\gl),
\end{eqnarray} 
In principle one would need to know the first order correction to $\left \bra \eta_i \right \ket_\gl$
to obtain (\ref{eq: derivee}) at the second order in $\gl$.

\medskip

\underline{\it For $\cA$ divergence free}, the term $ \bra \eta_i \ket_\gl$ disappears and the formula (\ref{eq: derivee 2}) simplifies
\begin{eqnarray}
\label{eq: dvp exp lambda}
\lim_{\tau \to \infty} {1 \over \tau} \log  \left \bra \exp \left( \gl  \bbQ^\cA(\tau)  \right) \right \ket
= \gl  \mean + {\gl^2 \over 2}  \var +  O(  {\gl^3}) \, .
\end{eqnarray}
where 
\begin{eqnarray*}
\mean = \lim_{\tau \to \infty} {\bra \bbQ^\cA(\tau)\ket \over \tau} =
 \sum_{i \in \CA} \cA_{(\bar i,i)} \ga_i \, .
\end{eqnarray*}
\begin{eqnarray}
\label{eq: variance free}
\var = \lim_{\tau \to \infty} {\bra \bbQ^\cA(\tau)^2 \ket_c \over \tau} 
= \sum_{(i,j) \in \cE_\gL}  (\cA_{(i,j)})^2  \left \bra \eta_i (1- \eta_j) \right \ket  
+ \sum_{i \in \CA} (\cA_{(\bar i,i)})^2  \left \bra \eta_i (1-\ga_i) + \ga_i (1 - \eta_i) \right \ket, 
\end{eqnarray}
where the sum is over all the oriented edges.
If there are no sources then the second term in (\ref{eq: variance free}) disappears.

\medskip

\underline{\it For a general field $\cA$}, we can use the decomposition (\ref{eq: decomposition}).
If $\CA \not = 0$, there is a representation of $V$ in terms of Green functions (\ref{eq: H}): for any $(i,j) \in \bar \gL$
\begin{eqnarray}
\label{eq: 3.17}
V_{(i,j)} = \cA_{(i,j)} + \sum_{k \in \gL}  {\rm div} \cA(k) \; \left( -G^{(k)}_i + G^{(k)}_j \right)\, .
\end{eqnarray}
Invariance (\ref{eq: gauge}) implies that the asymptotic formula for
(\ref{eq: 3.17}) is given by (\ref{eq: dvp exp lambda}) with $\cA$
replaced by $V$
\begin{eqnarray*}
\mean =  \sum_{i \in \CA} V_{(\bar i,i)} \ga_i \, .
\end{eqnarray*}
\begin{eqnarray}
\label{eq: variance explicite}
\var = \sum_{(i,j)\in \cE_\gL}  (V_{(i,j)})^2  \left \bra \eta_i (1- \eta_j) \right \ket  
+ \sum_{i \in \CA} (V_{(\bar i,i)})^2  \left \bra \eta_i (1-\ga_i) + \ga_i (1 - \eta_i) \right \ket
\, ,
\end{eqnarray}
where as before the sum is over all the oriented edges.

\medskip

Further simplications can be obtained if the system is in equilibrium at density $\bar \gr$, i.e. if the intensities of the sources are such that $\ga_i= \bar \gr, \  \gb_i = 1- \bar \gr$. 
In this case, $\bra \eta_i (1- \eta_j) \ket = \bar \gr (1-\bar \gr)$  and  by expanding \eqref{eq: variance explicite} with $V_b = \cA_b - \nabla_b H$, we get 
\begin{eqnarray*}
&&\var =  \bar \gr (1-\bar \gr)  \left[ \sum_{(i,j)\in \bar \cE_\gL}  
\cA_{(i,j)}^2  - 2  \sum_{(i,j)\in \bar \cE_\gL} \cA_{(i,j)}\; \nabla_{(i,j)} H 
+  \sum_{(i,j)\in \bar \cE_\gL} \left( \nabla_{(i,j)} H \right)^2  \right]\\
&& \qquad \qquad 
=\bar \gr (1-\bar \gr)  \left[ \sum_{(i,j)\in \bar \cE_\gL}  
\cA_{(i,j)}^2  + 4  \sum_{i\in \gL} {\rm div} \cA(i) \; H_i 
- 2 \sum_{i \in \gL}  \Delta H_i \; H_i 
\right]  \, ,
\end{eqnarray*}
where the second equation is obtained by summation by parts. 
From (\ref{eq: laplacien}), one has  $\Delta H_i ={\rm div} \cA (i)$ so that
\begin{eqnarray}
\label{eq: temporaire 1}
\var =  \bar \gr (1-\bar \gr)  \left[ \sum_{(i,j)\in \bar \cE_\gL}  
\cA_{(i,j)}^2  + 2  \sum_{i\in \gL} {\rm div} \cA(i) \; H_i \right]  \, .
\end{eqnarray}
Replacing $H$ by (\ref{eq: H}), we finally obtain
\begin{eqnarray}
\var =  \bar \gr (1-\bar \gr) \left[ \sum_{(i,j)\in \bar \cE_\gL}  
\cA_{(i,j)}^2  -   2 \sum_{i, k \in \gL}  G^{(k)}_i {\rm div} \cA(k) \;{\rm div} \cA(i) \right]\, ,
\label{eq: var ouvert}
\end{eqnarray}
where the sum is over all the oriented edges $(i,j)$.


\section{Two dimensional SSEP}
\label{sec: Z2}

In this section we will apply the general results of section 3 to the SSEP in the periodic square lattice $\gL = \{1,L\}^2$
with nearest neighbor jumps and derive explicit expressions in this case.
We consider the integrated current flowing through the edges in  
$\gG^\ell_L = \{L/2,L/2+1\} \times \{ 1,\ell \}$  given by 
\begin{eqnarray}
\label{eq: statio current discret}
Q^\ell(\tau) = \sum_{(i,i+\vec{e}_1) \in \gG^\ell_L} q_{(i,i+\vec{e}_1)}(\tau) 
-q_{(i+\vec{e}_1,i)}(\tau) \, .
\end{eqnarray}
with $\vec{e}_1=(1,0)$.
The integrated current $Q^\ell$ can be rewritten as $\bbQ^\cA$ defined in \eqref{eq: bbQ} with
\begin{eqnarray}
\label{eq: cA}
\forall i,j \in \gL, \qquad 
\cA_{i,j} = 
\begin{cases}
1, \quad {\rm if} \ (i,j) = (i,i + \ex) \in \gG^\ell_L \, ,\\
-1, \quad {\rm if} \ (i,j) = (i,i-\ex) \in \gG^\ell_L \, ,\\
0, \quad {\rm otherwise} \, .
\end{cases}
\end{eqnarray}

\begin{figure}[h]
\begin{center}
\leavevmode
\epsfysize = 5 cm
\psfrag{+}[cr]{$z^+_\ell$}
\psfrag{-}[cr]{$z^-_\ell$}
\psfrag{G}[cr]{$\gG^\ell_L$}
\psfrag{g}[cr]{$\gga$}
\epsfbox{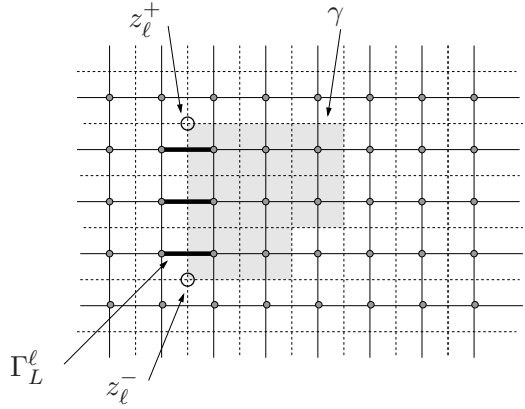}
\end{center}
\caption{The dashed lines represent the dual lattice and $\gG^\ell_L$ is depicted by the thick edges.
The function $H$ defined in equation (\ref{eq: cAH}) is equal to 1 in the grey region and 0 outside.}
\label{fig: dual}
\end{figure}

The gauge invariance (\ref{eq: gauge}) is easily illustrated in the 
two dimensional case. Let $z^+_\ell$ and $z^-_\ell$ be the 2 sites of the dual lattice 
such that $\gG^\ell_L$ is the set of edges intersected by the segment $(z^+_\ell,z^-_\ell)$ (see figure \ref{fig: dual}).
Let $\gga$ be another path connecting $z^+_\ell$ to $z^-_\ell$ on the dual lattice, then we can define 
the current $\bbQ^\cB(\tau)$ flowing through the edges intersecting $\gga$, where 
$\cB$ generalizes (\ref{eq: cA}) for the edges intersecting $\gga$. 
One can check that 
\begin{eqnarray}
\label{eq: cAH}
\cA = \cB  + \nabla H\, ,
\end{eqnarray}
for some $H$. Therefore, the statistics of the currents $\bbQ^\cA (\tau)$ and $\bbQ^\cB (\tau)$ are asymptotically the same at large times (\ref{eq: gauge}).

\bigskip

\subsection{Computation of the variance}

We turn now to the computation of 
\begin{eqnarray*}
\var = \lim_{\tau \to \infty} {\bra \bbQ^\ell (\tau)^2 \ket_c \over \tau} \, ,
\end{eqnarray*}
the asymptotic of the variance of $Q^\ell(\tau)=\bbQ^\cA(\tau)$ (\ref{eq: cA}) for large $\tau$.
On the periodic domain, the variance is given by (\ref{eq: variance explicite}) without the source term.
As the invariant measure is uniformly distributed, $\bra \eta_i (1-\eta_j) \ket$ depends only on the 
number $N$  of particles and  the size $L$.
Let $S_{L,N} = {N (L^2 -N) \over L^2 (L^2-1)} =\bra \eta_i (1- \eta_j) \ket$ for $i \not = j$, 
then the expression (\ref{eq: temporaire 1}) remains valid
\begin{eqnarray}
\label{eq: temporaire 2}
\var =  S_{L,N}   \left[ \sum_{(i,j)\in \cE_\gL}  
\cA_{(i,j)}^2  + 2  \sum_{i\in \gL} {\rm div} \cA(i) \; H_i \right]  \, ,
\end{eqnarray}
where $H$ is given by (\ref{eq: laplacien}) which reads now
\begin{eqnarray}
\label{eq: laplacien 2}
\forall i \in \gL, \qquad \Delta H_i = {\rm div} \cA (i) 
= 
\begin{cases} 
1, \qquad  & \text{if} \ i \in \gG^{\ell,+}_L = \{L/2,s\}_{1\leq s \leq \ell} \\
-1, \qquad & \text{if} \ i \in \gG^{\ell,-}_L = \{L/2 + \vec{e}_1 ,s\}_{1\leq s \leq \ell}\\
0, \qquad & \text{otherwise} 
\end{cases} 
\end{eqnarray}
Thus $H$ is equal to 
\begin{eqnarray}
\label{eq: H2}
H_i = - \sum_{k \in \gG^{\ell,+}_L}  G^{(k,k+\vec{e}_1)}_i \, ,
\end{eqnarray}
where the Green's function \eqref{eq: Green general} is replaced for any sites $k,k^\prime$ in $\gL$ by 
\begin{eqnarray}
\label{eq: Green torus}
\forall i \in \gL,\quad  \Delta G^{(k^\prime, k)}_i =
\gd_{i, k}- \gd_{i, k^\prime} \, .
\end{eqnarray}
From (\ref{eq: temporaire 2}), we finally obtain
\begin{eqnarray}
\label{eq: var periodic}
\var = 2  S_{L,N}  \left[ \ell  -   
\sum_{{i,k \in \gG^{\ell,+}_L }}  \left( G^{(k,k+\vec{e}_1)}_i  - G^{(k, k+\vec{e}_1)}_{i + \vec{e}_1} \right) \right]\, .
\end{eqnarray}

\medskip

The Green's function (\ref{eq: Green torus}) is given for any $i=(i_1,i_2)$ in $\gL$ by 
\begin{eqnarray}
\label{eq: Green torus exact}
G^{(k,k^\prime)}_i = \frac{1}{L^2} \sum_{q_1,q_2 \not=(0,0)} 
 \frac{\exp \big( {\bf i} \; q \cdot (i-k)  \big) - \exp \big( {\bf i} \; q \cdot (i-k^\prime) \big) }{4 -2 \cos(q_1) - 2 \cos(q_2)}    
\end{eqnarray}
where $ q \cdot j =q_1 j_1 + q_2 j_2$ stands for the scalar product
with $q_1 = 2 \pi {m_1 \over L}, q_2 = 2 \pi{m_2 \over L}$ for $m_1, m_2$ in $\{0,L-1\}$.
Thus (\ref{eq: var periodic}) becomes with the convention that ${1-\cos(q_2 \ell) \over 1-\cos(q_2 )} = \ell^2$ for $q_2 =0$
\begin{eqnarray}
\label{eq: V2 bis bis}
\var = 2 S_{L,N}  \left( \ell - {1 \over L^2} 
 \sum_{q_1,q_2 \not=(0,0)} {(1 -\cos(q_1)) ( 1-\cos(q_2 \ell) ) \over 
\big( 1 -\cos (q_2)\big) \big( 2 - \cos(q_1) -  \cos(q_2) \big) }\right) \, ,
\end{eqnarray}
where the subscript has been added to keep track of the dependence in $L,\ell$. 
Using the identity
\begin{eqnarray*}
{1 \over L}  \sum_{q_2} {1-\cos(q_2 \ell) \over 1-\cos(q_2 )} =
{1 \over L}  \sum_{m  =1}^{L-1} {1 -\cos \left( 2 \pi \ell {m \over L} \right) \over 
 1 -\cos \left( 2 \pi {m \over L} \right)  } + {\ell^2 \over L} = \ell \, ,
\end{eqnarray*}
we finally rewrite (\ref{eq: V2 bis bis}) as
\begin{eqnarray}
\label{eq: V2 bis}
\var =
2 S_{L,N}  \left( \frac{\ell^2}{ L^2} + \frac{1}{ L^2}
\sum_{q_1,q_2 \not=(0,0)} \frac{1- \cos \left(q_2 \ell  \right)}{2 -\cos(q_1) - \cos(q_2)} \right)
\end{eqnarray}
with $q_1 = 2 \pi {m_1 \over L}, q_2 = 2 \pi{m_2 \over L}$ for $m_1, m_2$ in $\{0,L-1\}$.

\bigskip

One can show, see Appendix B, that for large $L$ and $\ell$, 
expression (\ref{eq: V2 bis}) becomes for $h=\ell / L$
\begin{eqnarray}
 \var = {2 S_{L,N} \over \pi} \left[ \log L + h^2 + \log \left( { \sinh(\pi h) \over \pi} \right)
+ {3\log 2\over 2} + \gamma  +
\sum_{m \geq 1}   \log \left( 1 + { \sin^2 \big( \pi h \big) \over \sinh^2 (\pi m)}\right) \right]\, , 
\label{eq: asymptotic}
\end{eqnarray}
where $\gga \approx 0.577$ is the Euler constant.

\subsection{Time dependence}

To compare with the results of simulations, it is useful to calculate how the moments of $\bbQ^\cA (\tau)$ depend on $\tau$.  At finite times $\tau$, the gauge invariance (\ref{eq: gauge}) is of no use.
We focus now on systems with no sources and study the variance of $\bbQ^\cA (\tau)$ for a general field 
$\cA$ at finite time $\tau$.\\

Following the same procedure which led to (\ref{eq: derivee}),
(\ref{eq: derivee 2}), we get up to the second order in $\gl$,
\begin{eqnarray}
&& \partial_\tau  \left \bra \exp \left( \gl  \bbQ^\cA(\tau)  \right) \right \ket
= \gl \; \sum_{i \in \gL}  {\rm div} \cA (i) \; \left \bra \eta_i (\tau) \exp \left( \gl  \bbQ^\cA(\tau)  \right) \right \ket
\nonumber \\ 
&& \qquad \qquad \qquad   
+ {\gl^2 \over 2} \sum_{(i,j) \in \cE_\gL}  (\cA_{(i,j)})^2  \left \bra \eta_i (\tau) (1- \eta_j (\tau)) \exp \left( \gl  \bbQ^\cA(\tau)  \right) \right \ket   \, ,
\label{eq: Time dependence +}
\end{eqnarray}
where the second sum is over all the oriented edges.
We therefore need to determine $\bra \eta_i (\tau)  \exp \left(  \gl \bbQ^\cA(\tau)  \right) \ket$
to first order in $\gl$. To do so, we can write
\begin{eqnarray*}
&& \partial_\tau \left \bra \eta_i (\tau)  \exp \left(  \gl \bbQ^\cA(\tau)  \right) \right \ket
= 
\sum_{(k,j) \in \cE_\gL \atop k,j \not = i} (\exp( \gl \cA_{(k,j)}) - 1) \left \bra \eta_i (\tau) \eta_k (\tau) (1- \eta_j (\tau) ) \exp \left( \gl \bbQ^\cA(\tau)  \right) \right \ket\\ 
&& \qquad + \sum_{j \sim i } \exp( \gl \cA_{(j,i)}) \;  \left \bra \eta_j (\tau) (1- \eta_i (\tau) ) \exp \left( \gl \bbQ^\cA(\tau)  \right) \right \ket
- \left \bra \eta_i (\tau)  (1- \eta_j (\tau) ) \exp \left( \gl \bbQ^\cA(\tau)  \right) \right \ket \, ,
\end{eqnarray*}
where the first sum is over all the oriented edges $(k,j)$ which do not intersect $i$, but the second sum is over the neighbors $j$ of $i$.
As for \eqref{eq: derivee}, this expression can be derived by adding the contributions of all the single moves which may occur during the infinitesimal
time interval $d \tau$.
Expanding to first order in $\gl$, we get for a given $\tau$
\begin{eqnarray}
\label{eq: step 1}
&& \partial_\tau \left \bra \eta_i (\tau) \exp \left(  \gl \bbQ^\cA(\tau)  \right)  \right \ket
= \gl \sum_{(k,j) \in \cE_\gL \atop k,j \not = i} \cA_{ (k,j)}
\left \bra \eta_i  (\eta_k - \eta_j) \right \ket  \\
&& + \sum_{j \sim i}  \left \bra \eta_j(\tau)  \exp \left(  \gl \bbQ^\cA(\tau)  \right) \right \ket  - 
\left \bra \eta_i (\tau)  \exp \left(  \gl \bbQ^\cA(\tau)  \right) \right \ket  
+ \gl \cA_{(j,i)}  \left \bra \eta_j (1- \eta_i) \right \ket   \, . \nonumber
\end{eqnarray}
In the periodic case, 
the first term in (\ref{eq: step 1}) vanishes since
$\bra \eta_j (1- \eta_i) \ket = S_{L,N}$, is independent of $i,j$.  Hence
\begin{eqnarray}
&& \partial_\tau \left \bra \eta_i (\tau) \exp \left(  \gl \bbQ^\cA(\tau)  \right)  \right \ket
= \sum_{j; \  (i,j) \in \cE_\gL}   \left \bra \eta_j(\tau)  \exp \left(  \gl \bbQ^\cA(\tau)  \right) \right \ket  - 
\left \bra \eta_i (\tau)  \exp \left(  \gl \bbQ^\cA(\tau)  \right) \right \ket \nonumber  \\
&& \qquad \qquad + \gl S_{L,N} \sum_{j  \sim i } \cA_{(j,i)} \, . 
\label{eq: derivee densite +}
\end{eqnarray}

\medskip

We introduce for any site $k$ in $\gL$ the time dependent Green's function, solution of
\begin{eqnarray}
\label{eq: green time +}
\forall i \in \gL, \qquad 
\partial_\tau G^{(k)}_{\tau,i}  = \gD G^{(k)}_{\tau,i} + \gd_{i, k} =
 \sum_{j \sim i} \left(G^{(k)}_{\tau,j} - G^{(k)}_{\tau,i} \right) +
\gd_{i, k} \, ,
\end{eqnarray}
with the initial condition $G^{(k)}_{0,i}  =0$.
Integrating (\ref{eq: derivee densite +}), one obtains to first order in $\gl$
\begin{eqnarray}
\label{eq: eta 1er ordre +}
\left \bra \eta_i (\tau)  \exp \left(  \gl \bbQ^\cA(\tau)  \right) \right \ket
= \left \bra \eta_i \right \ket + \gl S_{L,N} \sum_k G^{(k)}_{\tau,i}
\sum_{j \sim k}  \cA_{(j,k)}  
= \left \bra \eta_i \right \ket - \gl S_{L,N} \sum_k G^{(k)}_{\tau,i}  {\rm div} \cA (k) \, .  
\end{eqnarray}
Using (\ref{eq: eta 1er ordre +}) in (\ref{eq: Time dependence +}), we
get for the second order term in $\gl$ 
\begin{eqnarray}
\label{eq: Time dependence 2 +}
&& \partial_\tau \left \bra \exp \left( \gl  \bbQ^\cA(\tau)  \right) \right \ket =
 \gl^2 S_{L,N} \; \left( {1 \over 2} \sum_{(i,j) \in \cE_\gL}  (\cA_{(i,j)})^2   
-  \sum_{k \in \gL \atop i \in \gL} G^{(k)}_{\tau,i} \; {\rm div} \cA (k) \; {\rm div} \cA (i) \right) 
\, ,
\end{eqnarray}
where the first sum is over all the oriented edges.

\bigskip

For the  periodic square lattice $\gL = \{1,L\}^2$,
the Green's function (\ref{eq: green time +}) is given for any $k = (k_1,k_2)$ and 
$i = (i_1,i_2)$ in $\gL$ 
\begin{eqnarray*}
G^{(k)}_{\tau,i} = {1 \over L^2}
\left[ \tau + \sum_{q_1,q_2 \not=(0,0)} 
 \frac{1 - {\rm e}^{- (4 -2 \cos(q_1) - 2 \cos(q_2)) \tau }}{4 -2 \cos(q_1) - 2 \cos(q_2)}   
\exp \left( {\bf i} \; q \cdot (i-k) \right)  \right] \, ,
\end{eqnarray*}
with $q_1 = 2 \pi {m_1 \over L}, q_2 = 2 \pi{m_2 \over L}$ for $m_1, m_2$ in $\{0,L-1\}$.
Using (\ref{eq: Time dependence 2 +}), we deduce an exact expression
for the variance of the current $Q^\ell(\tau)$ through a slit ($\cA$ is given by (\ref{eq: cA}))
\begin{eqnarray*}
\bra \bbQ^\ell (\tau)^2 \ket_c = 2\, S_{L,N} \int_0^\tau  d s \, 
\left({\ell}-{1 \over L^2} 
\sum_{q_1,q_2 \not=(0,0)}\frac{(1 -{\rm e}^{-(4 -2 \cos(q_1)- 2 \cos(q_2)) s } ) }{2- \cos(q_1)- \cos(q_2)} {(1-\cos(q_1)) (1-\cos(q_2 \ell))\over 1-\cos (q_2)}\right) \, . 
\end{eqnarray*}
We finally get
\begin{eqnarray}
\label{eq: var t}
{\bra \bbQ^\ell (\tau)^2 \ket_c \over \tau}
=\var + {S_{L,N} \over \tau L^2} 
\sum_{q_1,q_2 \not=(0,0)}\frac{(1 -{\rm e}^{-(4 -2 \cos(q_1)- 2 \cos(q_2)) \tau } ) }{(2- \cos(q_1)- \cos(q_2))^2} {(1-\cos(q_1)) (1-\cos(q_2 \ell))\over 1-\cos (q_2)}\, ,
\end{eqnarray}
where $\var$, given in (\ref{eq: V2 bis bis}) or (\ref{eq: V2 bis}), is related to the large $\tau$ asymptotics.

\subsection{Numerical simulations}

\begin{figure}[h]
\begin{center}
\leavevmode
\epsfysize = 5 cm
\epsfbox{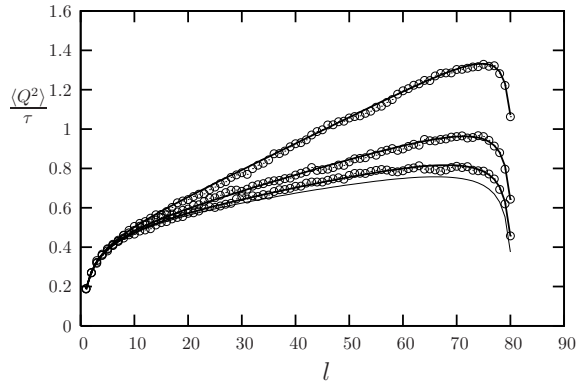}
\end{center}
\caption{${\bra Q^\ell(\tau)^2 \ket \over \tau}$ is measured versus $\ell$ in numerical simulations for the SSEP on a square of $80 \times 80$ sites  for times $\tau= 250, 750, 2500$
(the results decrease with $\tau$).
 The continuous lines represent the theoretical predictions (\ref{eq: var t}) at these times, as well as the limit $\tau = \infty$ (\ref{eq: V2 bis}). Expression (\ref{eq: var t}) fully agrees with the simulations.}
\label{fig: figdata}
\end{figure}

We show now the results of the simulations of the SSEP on a square lattice of size $L=80$ with periodic boundary conditions at density $\bar \gr = 1/4$ (without reservoirs). The initial condition is chosen at equilibrium (i.e. the $L^2 \bar \gr$ particles are put at random positions on the square lattice). 
For each simulation, we measured the flux $Q^\ell(\tau)$ through a slit of microscopic length $\ell$
during time $\tau$ (see figure \ref{fig: reservoirs} and (\ref{eq: statio current discret})).

In figure \ref{fig: figdata}, our data for ${\bra Q^\ell(\tau)^2 \ket \over \tau}$
are compared with the predictions obtained from (\ref{eq: var t}) for different times 
$\tau= 250, 750, 2500$. The simulations are averaged over $10^4$ realizations. 
We see that unless the time is long enough, the results differ significantly from their infinite time 
limit (\ref{eq: V2 bis}).
One can notice that for short times, the variance grows essentially linearly wrt $\ell$ as the current fluctuations are simply the sum of the (almost) independent contributions of the local current fluctuations along the slit. 

\begin{figure}[h]
\begin{center}
\leavevmode
\epsfysize = 5 cm
\epsfbox{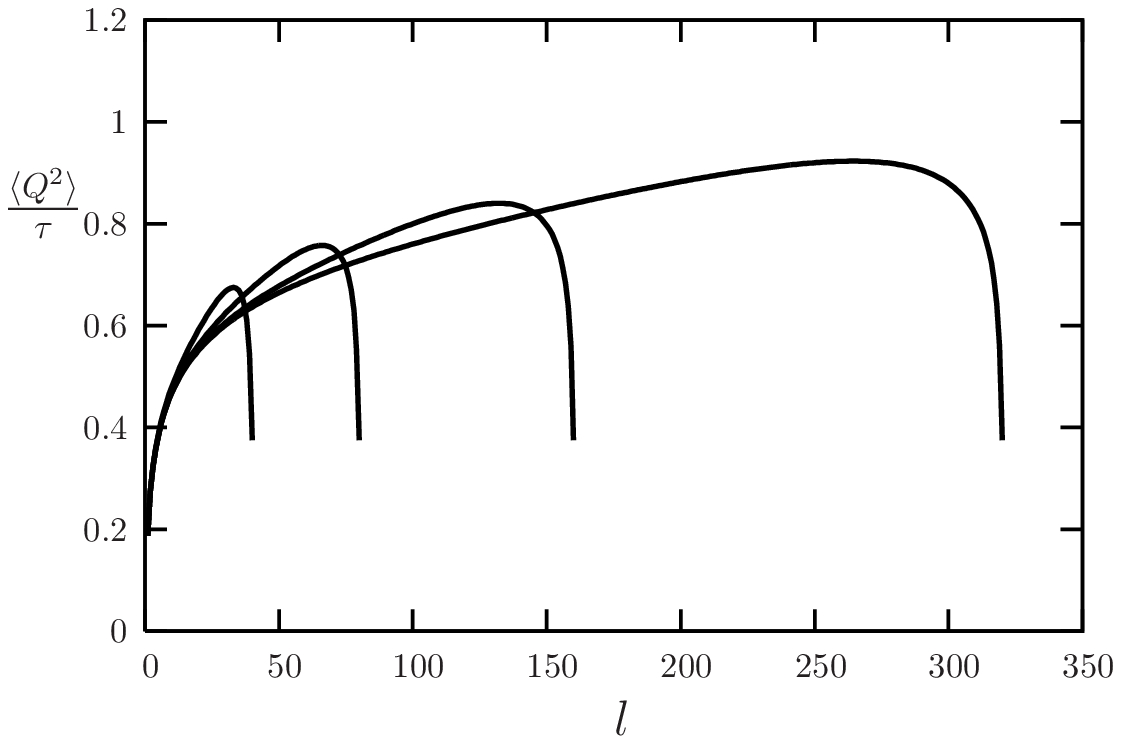}
\end{center}
\caption{Theoretical prediction (\ref{eq: V2 bis}) of $\lim_{\tau \to \infty} {\bra Q^\ell(\tau)^2 \ket \over \tau}$ versus $\ell$ for $L=40,80,160,320$.}
\label{fig: figth1}
\end{figure}

In figure \ref{fig: figth1}, the theoretical curve  $\var = \lim_{\tau \to \infty} {\bra Q^\ell(\tau)^2 \ket \over \tau}$  computed in (\ref{eq: V2 bis}) is shown for several system sizes $L=40,80,160,320$.
One can notes that the variance of the current flowing through the whole system ($\ell=L$) is independent of $L$.

\begin{figure}[h]
\begin{center}
\leavevmode
\epsfysize = 5 cm
\epsfbox{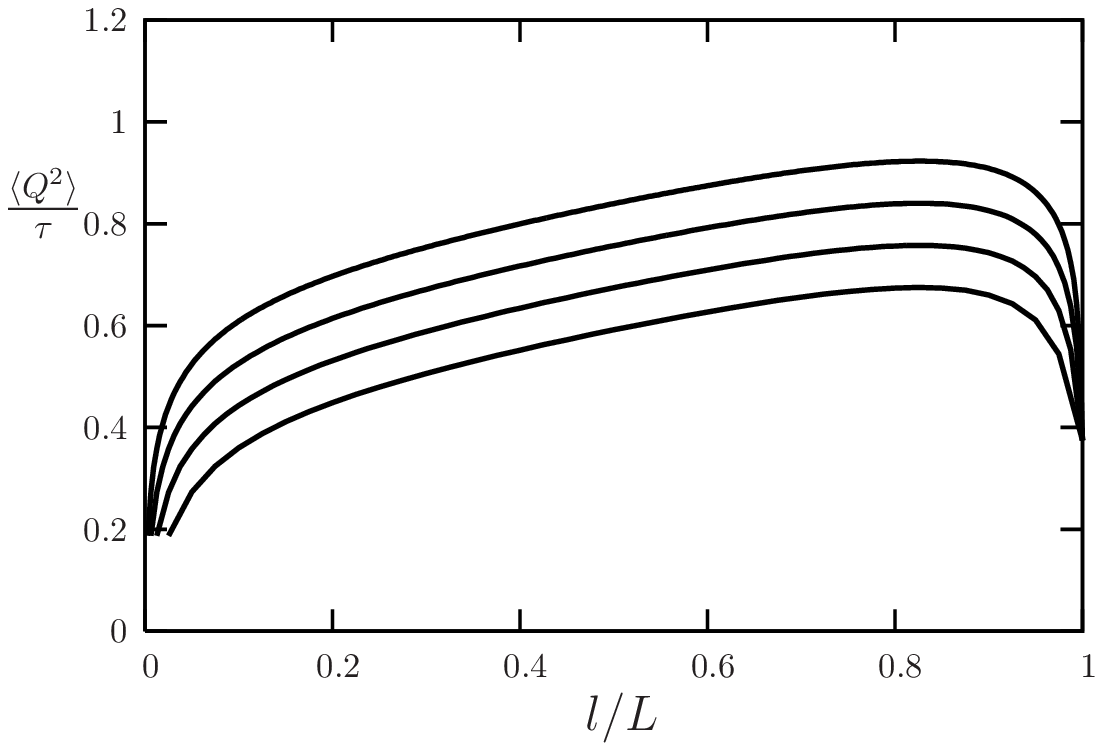}
\end{center}
\caption{Theoretical prediction (\ref{eq: V2 bis}) of $\lim_{\tau \to \infty} {\bra Q^\ell(\tau)^2 \ket \over \tau}$ versus $\ell / L$ for $L=40,80,160,320$.}
\label{fig: figth2}
\end{figure}

In figure \ref{fig: figth2}, the same data as in figure \ref{fig: figth1} are shown 
but the horizontal axis is now $\ell / L$. One can see that for large $L$, $\var$ grows
linearly with $\log L$ as predicted in (\ref{eq: asymptotic}).

	\section{Conclusion}

 In this paper, we have computed the variance of the local current for the symmetric simple exclusion process on general graphs 
with reservoirs \eqref{eq: variance explicite} or without \eqref{eq: var ouvert}.
In two dimensions, our exact expression leads to the asymptotics of the variance through a slit (\ref{eq: asymptotic}).
The logarithmic dependence of the variance confirms that vortices dominate the local current 
fluctuations. As a consequence a fluctuation of the partial current, say the current flowing through 99\% of the system, does not obey the same scaling as a fluctuation of the total current.
For two dimensional diffusive models, we have also seen that the hydrodynamic large deviation theory
does not catch the correct scaling of the current deviations \eqref{G0}.
Finite time corrections to the variance were also computed \eqref{eq: var t} and compared to numerical data (figure \ref{fig: figdata}).
Finally the fluctuation relation (\ref{ft}) for partial currents is discussed in Appendix A.

It would be interesting to investigate the scaling of partial current deviations in higher dimensions. Another challenging issue is the computation of the full large deviation functional for partial currents.

\section*{Appendix A: The fluctuation theorem and partial currents}


For the total current, the fluctuation theorem (\ref{ft}) holds (in any dimension)
\begin{equation}
\label{eq: ft2}
G_{L,L}(\jj;\rho_a,\rho_b) - G_{L,L} (-\jj;\rho_a,\rho_b)= \jj [ \log z(\rho_b) - \log z(\rho_a)] \, .
\tag{A.1}
\end {equation}
This fluctuation relation is based on a global symmetry: the fluctuation to produce the current $-J$
is simply the time reversal of the fluctuation to produce the current $J$.
One may wonder how this generalizes to the function $G_{L,\ell}$.
In this appendix, we show by considering a very simplified model that the  fluctuation relation (\ref{eq: ft2}) is in general not satisfied for  partial current deviations.

\medskip

\begin{figure}[h]
\begin{center}
\leavevmode
\epsfysize = 3 cm
\psfrag{b}[T]{$\gga$}
\psfrag{a}[Br]{$\ga$}
\psfrag{c}[Br]{$\gd$}
\psfrag{d}[T]{$\gb$}
\psfrag{j}[T]{$Q^\prime_\tau$}
\epsfbox{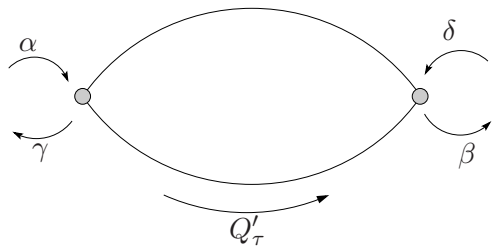}
\end{center}
\caption{A reservoir at density $\gr_a$ (resp $\gr_b$) is acting on the left (resp right) site by creating particles at rate $\ga$ (resp $\gd$) and annihilating particles at rate $\gga$ (resp $\gb$).
We consider the large deviations of the current $Q^\prime_\tau$ flowing through the lower edge.}
\label{fig: toy}
\end{figure}

We consider the SSEP with two sites $\{1,2\}$ connected to reservoirs.
At site 1, creation (resp annihilation) occurs at rate $\ga$ (resp $\gga$) and  at site 2, creation (resp annihilation) 
occurs at rate $\gd$ (resp $\gb$) (see figure \ref{fig: toy}).
The exchanges between sites $\{1,2\}$ obey the usual exclusion rule, but they can 
occur through two edges with rate 1. On the one hand the model behaves like a SSEP
with exchange rate 2. Thus the total current flowing from site 1 to site 2
obeys the fluctuation relation \eqref{eq: ft2} (with $\gr_a = {\ga \over \ga + \gga}$ and 
$\gr_b = {\gd \over \gd + \gb}$). 
On the other hand one can also consider a current deviation $J$ through one of the two edges.
Heuristically one can see that the system is going to use different 
strategies to produce a current $J$ or $-J$. Imagine the extreme case 
with only creation at site 1 and annihilation at site 2 ($\gga=0$ and $\gd=0$).
For the total current, there is no way of producing a negative flux and the 
relation \eqref{eq: ft2} is degenerate: $\log z(\rho_b) - \log z(\rho_a)=-\infty$. 
On the other hand, a negative current can be achieved through the lower edge by letting 
a single particle cross the lower edge from site 2 to site 1 and then use 
the upper edge to go back to site 2.
This latter mechanism mimics the vortices discussed in section \ref{sec: vortices}.
Thus, the fluctuations to produce current deviations $J$ or $-J$ are not related by time reversal.
In general, both mechanisms (total current deviation and local vortices) combine and there is no reason to expect a symmetry such as
\eqref{eq: ft2}.

\bigskip

We analyze now the toy model analytically.
We define $Q^\prime_\tau$ as the integrated current flowing through the lower edge 
during the time interval $[0,\tau]$  (see figure \ref{fig: toy}). 
Instead of trying to check an expression like \eqref{eq: ft2} for the large deviation function 
we look for a symmetry at the level of its Legendre transform.
As in \cite{DDR}, one knows that 
\begin{eqnarray*}
\forall \gl, \qquad 
\lim_{\tau \to \infty} \frac{1}{\tau} \log \left \bra \exp \big( \gl Q^\prime_\tau \big) \right \ket 
= \mu (\gl) \, ,
\end{eqnarray*}
where $\mu (\gl)$ is the largest eigenvalue of the operator
\begin{eqnarray*}
L_\gl = \left(
\begin{array}{llll}
 -\ga-\gd & \gga & \gb & 0 \\
 \ga & -\gga-\gd-2 & 1+e^{-\gl} & \gb \\
 \gd & 1+e^{\gl} & -\ga- \gb -2 & \gga \\
 0 & \gd & \ga & -\gga -\gb
\end{array}
\right)
\end{eqnarray*}
The fluctuation relation \eqref{eq: ft2} would say that there exists a constant $E$ such that
\begin{equation}
\label{eq: legendre}
\forall \gl, \qquad  \mu(-\gl-2E) = \mu(\gl)
\tag{A.2}
\end{equation}

In order to prove that the previous relation does not hold, 
we consider for simplicity the case $\ga=2,\gga=1$ and $\gd=1,\gb=2$. 
Then the characteristic polynomial of $L_\gl$ is
\begin{eqnarray*}
P(u) = (3 + u) (16 + u (44 + u (13 + u)) - 2 (8 + u) \cosh[\gl] - 6 \sinh[\gl])
\end{eqnarray*}
For (\ref{eq: legendre}) to be satisfied, $\mu(\gl)$ should be a root of this polynomial and 
of the polynomial associated to $L_{-\gl-2E}$. This implies that 
\begin{eqnarray*}
2 (8 + \mu(\gl)) \cosh[\gl] + 6 \sinh[\gl]
= 2 (8 + \mu(\gl)) \cosh[-\gl-2E] + 6 \sinh[-\gl-2E]
\end{eqnarray*}
leading to 
\begin{eqnarray*}
\forall \gl, \qquad \mu(\gl) = - 3\coth [E] -8 \quad \text{or}  \qquad \mu(\gl)= -3\, .
\end{eqnarray*}
As $\mu(\gl)$ cannot be independent of $\gl$, we obtained a contradiction. Thus
the fluctuation relation \eqref{eq: legendre} does not hold in this toy 
model.

For the total current a similar calculation shows that $\mu(\gl)$ is the root of
\begin{eqnarray*}
Q(u) =(3 + u) (20 + u (6 + u) (7 + u) - 20 \cosh[\gl] - 12 \sinh[\gl]) \, .
\end{eqnarray*}
This is invariant under the symmetry $\gl \to - 2 \log 2 - \gl$, implying that 
(\ref{eq: legendre}) is satisfied.

\section*{Appendix B: }

In this appendix we derive the large $L,\ell$ expression (\ref{eq: asymptotic}) of the
variance  (\ref{eq: V2 bis}).

\bigskip

Define $I_N$ and $J_N$ by
$$I_N= \sum_{n=1}^N {1 \over n^2 + b^2} 
\qquad {\rm and} \qquad
J_N=  \int_0^N {dx  \over b^2 + x^2}$$
One has for large $N$
\begin{eqnarray}
I_N = J_N + \pi^2 \left[ {1 \over 2 \pi b \tanh( \pi b) } - {1 \over 2
\pi^2 b^2} - {1 \over 2 \pi b} \right] + o(1) = J_N + \pi^2 \left[ {1 \over \pi b [\exp(2 \pi b) -1] } - {1 \over 2
\pi^2 b^2}  \right] + o(1)
\label{eq: approx n^2}
\end{eqnarray}
Recall also that 
\begin{eqnarray}
\int_0^1 {d x \over 2 - B - \cos(2 \pi x)}= {1 \over \sqrt{(1-B)(3-B)}} \  .
\label{eq: approx integrale}
\end{eqnarray}

\bigskip

From (\ref{eq: approx n^2}, \ref{eq: approx integrale}), one can show, by taking $b^2 = L^2(1- \cos q_2)/(2 \pi^2)$, that for $0 < q_2 < 2 \pi$
\begin{eqnarray}
 {1 \over L} \sum_{n_1=0}^{L-1} {1 \over 2 - \cos q_2 - \cos{2 \pi n_1
\over L}} &=&
 {1 \over \sqrt{(1 - \cos q_2)(3-\cos q_2) }} \nonumber \\
&& \qquad \qquad 
+  {1 \over    \sin ({q_2 \over 2}) (\exp [ 2 L \sin ({q_2 \over 2})] -1) } + o(1) \, .
\label{eq: approx n^2 bis}
\end{eqnarray}
The main contribution to the difference between the sum in \eqref{eq: approx n^2 bis} and 
integral (\ref{eq: approx integrale}) is 
given by the terms with $n_1$ close to 0 or to $L$. In both cases $(1- \cos{2 \pi n_1
\over L})$ can approximated by its second order expansion and 
the last term in \eqref{eq: approx n^2 bis} is obtained thanks to \eqref{eq: approx n^2}.

\bigskip

\eqref{eq: approx n^2 bis} can be rewritten as
$${1 \over L} \sum_{n_1=0}^{L-1} {1 \over 2 - \cos q_2 - \cos{2 \pi n_1
\over L}} =
 {1  - \sqrt{3- \cos q_2 \over 2} \over \sqrt{(1 - \cos q_2)(3-\cos
q_2) }} +  {1 \over 2 \sin {q_2 \over 2}}+
 {1 \over    \sin ({q_2 \over 2}) (\exp [ 2 L \sin ({q_2 \over 2})] -1) } 
+ o(1)$$
One can then perform the sum over $q_2$. For large $L$, the first term
becomes an integral
$$\int_0^1 dx {1 - \sqrt{3- \cos(2 \pi x) \over 2} \over \sqrt{[1- \cos(
2 \pi x)][3-\cos(2 \pi x)]}} = {1 \over 2 \pi} \int_0^{\pi} \, d \phi \, {1 - \sqrt{1 +
\sin^2 \phi} \over \sin \phi  \  \sqrt{1 +
\sin^2 \phi}}= - { \log 2 \over  2 \pi }$$
For large $L$ one can also show that
$${1 \over L} \sum_{n=1}^{L-1} {1 \over 2 \sin{n \pi \over L}} \simeq {1 \over \pi}
\left[ \log L + \log \left({2 \over \pi} \right) + \gamma_E  \right] + o( 1 )$$
For large $l$ and $L$ with $l = L h$
$${1 \over L} \sum_{n=1}^{L-1} { \cos( {2n  \pi \over L} l ) \over 2 \sin{n
\pi \over L}} \simeq - {1 \over 2 \pi} \log \left( 2 - 2 \cos (2 \pi  h)
 \right)  = - {1 \over \pi} \log (2 \sin(\pi h))$$
There is also the identity
$$\sum_{n=1}^\infty  {2 [1 - \cos (2 \pi n h)]\over n \pi ( e^{2 n
\pi} -1) } =  {1 \over \pi } \sum_{m=1}^\infty \log \left( 1 + {\sin^2 (\pi h) \over
\sinh^2(\pi m)} \right) 
$$
 Putting everything together one gets that
\begin{eqnarray}
 \nonumber 
&& {1 \over L^2} \sum_{n_1=0}^{L-1}  \sum_{n_2=1}^{L-1} {1  - \cos(q_2 l) \over 2 - \cos q_2 - \cos q_1
}  \simeq \\  && {1 \over \pi} \left[ \log L + \log \left( {\sin (\pi
h) \over \pi} \right) + {3  \log 2 \over 2} + \gamma_E + \sum_{m \geq 1}
\log \left( 1 + {\sin^2(\pi h) \over \sinh^2   ( \pi m) } \right) \right]
 \nonumber 
\end{eqnarray}
 Note that for $h $ small,  i.e for $1 \ll l \ll L$, one recovers a well
known expression (see \cite{ID} page 198).

\bigskip

\noindent
{\it Acknowledgements:}
The research was supported in part by NSF Grant DMR01-279-26 and AFOSR Grant AFFA9550-04.
T.B. and B.D. acknowledge the support of the ANR {\it LHMSHE}.


\begin{thebibliography}{99}

%
%
%

\bibitem{gaspard} 
D. Andrieux and P. Gaspard,
Network and thermodynamic conditions for a single macroscopic current fluctuation theorem,
{\it Comptes rendus Physique}, {\bf 8} No 5-6, 579--590 (2007)


\bibitem{BDGJL4}
L.  Bertini, A. De Sole, D. Gabrielli, G. Jona-Lasinio, C. Landim,
Non equilibrium current fluctuations in stochastic lattice gases,  
{\it J. Stat. Phys.}  {\bf 123},  no. 2, 237--276  (2006)

\bibitem{BDGJL5} L.  Bertini, A.  De Sole, D.  Gabrielli, G.
Jona--Lasinio, C.  Landim, Current Fluctuations in Stochastic Lattice Gases,
{\it Phys.  Rev.  Lett.}~{\bf 94}, 030601 (2005)

\bibitem{BDGJL6} L.  Bertini, A.  De Sole, D.  Gabrielli, G.
Jona--Lasinio, C.  Landim, 
Large deviations of the empirical current in interacting particle systems,
{\it Theory of Probability and its Applications} {\bf 51} No 1, 2--27, (2007)

\bibitem{BD1}
T.  Bodineau, B. Derrida,
 Current fluctuations in nonequilibrium diffusive systems: An additivity
principle {\it Phys. Rev. Lett.} {\bf 92},  180601 (2004)


\bibitem{BD2}
T.  Bodineau, B. Derrida,
Distribution of current in nonequilibrium diffusive systems and phase transitions,
{\it Phys. Rev. E} (3)  {\bf 72},  no. 6, 066110   (2005)

\bibitem{BD4}
T.  Bodineau, B. Derrida,
Cumulants and large deviations of the current through non-equilibrium steady states,
{\it Comptes rendus Physique}, {\bf 8} No 5-6, 540--555 (2007)


\bibitem{BLR} F. Bonetto, J.L. Lebowitz, L. Rey-Bellet,
Fourier's law: a challenge to theorists, Mathematical Physics 2000,
128-150,  {\it Imperial College Press} (2000).  math-ph/0002052


\bibitem{BL}
F. Bonetto, J. Lebowitz, 
Thermodynamic Entropy Production Fluctuation in a Two Dimensional
Shear Flow Model, {\it Physical Review E}, 64, 056129-1 to 056129-9, (2001)

\bibitem{Ciliberto}
S. Ciliberto, C. Laroche,
An experimental test of the Gallavotti-Cohen fluctuation theorem, {\it Journal de Physique} IV, Vol.8 , 215 (1998)



\bibitem{DDR}
 B. Derrida, B. Dou{\c c}ot, P.-E. Roche,  
 Current fluctuations in the one dimensional Symmetric Exclusion Process with open boundaries, 
{\it J. Stat. Phys.} 115, 717-748 (2004)

\bibitem{DV}
M. Donsker, S.R.S. Varadhan, Large deviations from a hydrodynamic scaling limit.  {\it Comm. Pure Appl. Math.}  {\bf 42},  no. 3, 243--270  (1989)


\bibitem{ECM} 
D.J. Evans, E.G.D. Cohen, G.P. Morriss, 
Probability of second law violations  in shearing steady states,
{\it Phys. Rev. Letts.} {\bf 71}, 2401 (1993)

\bibitem{ES} 
D.J. Evans, D.J. Searles, The Fluctuation Theorem, {\it Advances in Physics} {\bf 51}, 1529-1585 (2002)

\bibitem{GC}    
G. Gallavotti, E.D.G. Cohen, Dynamical ensembles in stationary states, {\it J. Stat. Phys.} {\bf 80}, 931-970 (1995)

\bibitem{G2}    
G. Gallavotti,
Entropy production in nonequilibrium thermodynamics: a point of view, {\it Chaos} {\bf 14}, 680--690, (2004)



\bibitem{ID} 
C. Itzykson, J.M. Drouffe, {\em Statistical field theory. Vol. 1. From Brownian motion to renormalization and lattice gauge theory}. Cambridge Monographs on Mathematical Physics (1989). 

\bibitem{KL} C. Kipnis, C. Landim,
{\em Scaling limits of interacting particle systems}, Springer (1999)

\bibitem{KOV} C. Kipnis, S. Olla, S. Varadhan, Hydrodynamics and
large deviations for simple exclusion processes, {\it Commun. Pure
Appl. Math.} {\bf 42}, 115-137 (1989)

\bibitem{K}
J. Kurchan, Fluctuation Theorem for stochastic dynamics,  {\it J.  Phys.} {\bf A31} 3719,  (1998)

\bibitem{LS}   
J.L. Lebowitz, H.~Spohn, A Gallavotti-Cohen Type Symmetry
in the Large Deviation Functional for Stochastic Dynamics
{\it J.   Stat. Phys.} {\bf 95}, 333-366 (1999)


\bibitem{LLP} S. Lepri, R. Livi, A. Politi,
Thermal conduction in classical low-dimensional lattices, {\it  Phys. Rep.} {\bf    377},  1-80 (2003)

\bibitem{M1}
C. Maes, The fluctuation theorem as a Gibbs property, {\it J. Stat. Phys.} {\bf 95}, 367-392  (1999)

\bibitem{spohn} H. Spohn,
{\em Large scale dynamics of interacting particles}, Springer (1991)




\end{thebibliography}
\end{document}